\documentclass[sigconf,nonacm]{acmart}

\usepackage{color}
\usepackage{multirow}
\usepackage{subfig}
\usepackage{xcolor}
\usepackage{lipsum}
\usepackage{listings}
\usepackage{glsllst} 
\usepackage{rmlst}   

\title[A Supervised Learning Framework for Network-Aware Scheduling of Data-Intensive Workloads]{Learning to Schedule: A Supervised Learning Framework for Network-Aware Scheduling of Data-Intensive Workloads}

\author{Sankalpa Timilsina}
\affiliation{
    \institution{Tennessee Technological University}
    \city{Cookeville}
    \state{TN}
    \country{USA}
}
\email{stimilsin43@tntech.edu}

\author{Susmit Shannigrahi}
\affiliation{
    \institution{Tennessee Technological University}
    \city{Cookeville}
    \state{TN}
    \country{USA}
}
\email{sshannigrahi@tntech.edu}



\begin{document}

\begin{abstract}
Distributed cloud environments hosting data-intensive applications often experience slowdowns due to network congestion, asymmetric bandwidth, and inter-node data shuffling. These factors are typically not captured by traditional host-level metrics like CPU or memory. Scheduling without accounting for these conditions can lead to poor placement decisions, longer data transfers, and suboptimal job performance. We present a network-aware job scheduler that uses supervised learning to predict the completion time of candidate jobs. Our system introduces a prediction-and-ranking mechanism that collects real-time telemetry from all nodes, uses a trained supervised model to estimate job duration per node, and ranks them to select the best placement. We evaluate the scheduler on a geo-distributed Kubernetes cluster deployed on the FABRIC testbed by running network-intensive Spark workloads. Compared to the default Kubernetes scheduler, which makes placement decisions based on current resource availability alone, our proposed supervised scheduler achieved 34–54\% higher accuracy in selecting optimal nodes for job placement. The novelty of our work lies in the
demonstration of supervised learning for real-time, network-aware job scheduling on a multi-site cluster.
\end{abstract}

\maketitle


\section{Introduction} \label{sec:introduction}
Scheduling is the process of assigning tasks or jobs to resources over time. An efficient scheduler improves system utilization, reduces job latency, and balances load across nodes. The rise of data-intensive applications across science, analytics, and AI/ML has led to growing demand for scalable, distributed infrastructure \cite{hey2009fourth,mehran2023scheduling}. These workloads generate and process massive datasets, making them sensitive to how jobs are scheduled. Poor or random scheduling decisions can lead to increased job completion times, congested network paths, and under- or over-utilized nodes \cite{marchese2025enhancing}. Informed scheduling is key to performance. For instance, if a Spark shuffle-heavy job, where large volumes of intermediate data must be exchanged across nodes, is scheduled on a node experiencing high outbound network traffic, it may significantly delay subsequent stages, even if the node has idle CPU cores.

Recent works have explored reinforcement learning (RL) and heuristic-based scheduling for telemetry-aware orchestration \cite{chowdhury2019drls,marchese2022communication}. However, RL methods are often sample-inefficient, require reward engineering, and are difficult to interpret \cite{glanois2024survey}. Heuristics, while simpler, may not generalize across workload types or cluster states \cite{senjab2023survey}.

In this work, we propose a network-aware, supervised learning–based scheduler that predicts job completion time per node using real-time cluster telemetry. Our methodology is modular, generic, and should be broadly applicable across diverse cluster and workload environments. Unlike RL-based approaches, our method does not require online exploration. And unlike static heuristics, it is trained on real executions, which allows it to learn how both current network conditions and resource load affect job runtime, and adapt accordingly. 

The scheduler collects live metrics such as inbound and outbound network traffic volume for each node, using telemetry data exposed by node-level exporters. In addition, it gathers host-level metrics such as CPU and memory usage to avoid partial observability \cite{michael2010partial}. The framework uses this live telemetry to predict job completion time across all nodes and schedules each job to the one with the shortest predicted duration. Our contributions are:
\begin{itemize}
    \item We design a network-aware scheduling system that uses supervised learning to predict job completion time per node from real-time telemetry and job configuration to guide placement decisions.
    \item We present a preliminary evaluation of our framework on the FABRIC testbed by deploying data-intensive workloads on a geo-distributed Kubernetes cluster spanning multiple sites.
    \item We show that our approach improves node selection accuracy by 34–54\% compared to the default Kubernetes scheduler for job placement.
\end{itemize}

\section{Motivation} \label{sec:motivation}
We establish our motivation based on the need for network-awareness in data-intensive applications, our choice of Kubernetes and Spark as representative platforms, the rationale for using supervised learning, and considerations around deployability and adaptability of our approach in production environments.

\subsection{Network-awareness for Data-Intensive Workloads}
Data-intensive applications involve massive data shuffles and I/O operations that make the network a critical performance factor, often more so than compute \cite{he2016firebird}. For example, in a network-intensive large-scale data-processing engine like Apache Spark, datasets are often cached in memory, which in turn makes network throughput the limiting resource for shuffles and distributed operations \cite{ULLAH2024103837}. Unbalanced data locality or poor placement of tasks can lead to severe network congestion and degrade cluster performance \cite{marchese2022communication}. Thus, network-aware scheduling policies that account for real-time network conditions are needed to avoid unnecessary data transfers and prevent network bottlenecks from undermining application performance.

\subsection{Insurgence of Cloud-Native Applications}
Cloud-native applications are highly distributed (often composed of many microservices or tasks) and are generally network-intensive in their interactions \cite{saleh2022empirical}. These systems have seen widespread adoption—by 2024, over 89\% of organizations had adopted cloud-native technologies, with Kubernetes remaining the dominant orchestration platform at over 93\% usage \cite{CNCF_2025}. At the same time, frameworks like Apache Spark are increasingly deployed in such environments to support large-scale data processing.

We select Kubernetes as the orchestration platform and Apache Spark as the representative data-intensive application framework to ensure our solution is relevant to modern cloud-native big data deployments.

\subsection{Supervised Learning for Scheduling}
Prior efforts like heuristic systems often require extensive tuning to specific workload characteristics, and their one-size-fits-all logic struggles when faced with dynamic or unforeseen scenarios. Another issue is that heuristics are inherently brittle and myopic as they lack the ability to learn and improve. The effectiveness of such systems is often limited in real-world use due to dynamic workload variations and changing system configurations \cite{senjab2023survey}.  

RL-based systems typically require a huge number of trial runs to converge on a good scheduling policy, because they learn only from incremental rewards. This is computationally expensive and slow \cite{han2023survey}. Another concern is stability and convergence guarantees. RL algorithms can be unstable i.e., small changes in hyperparameters or reward design can lead to very different outcomes. They often converge to local optima. This is RL’s poor generalization, where the policy overfits to the training scenario unless a very broad range of environments is used in training \cite{torres2023encyclopedia_drl_manufacturing}.

Supervised models generalize patterns from historical data and are easier to update when workload characteristics change. They can capture complex nonlinear relationships (e.g., between job attributes and optimal scheduling decisions) that are generally hard to encode in static heuristics \cite{delimitrou2014quasar}. Unlike RL, supervised learning does not require an active simulation loop to improve and can utilize existing logs and off-policy data to train offline, so there’s no risk of disrupting live systems during learning.

We would like to make a note that the choice is not always binary. We use supervised learning because it is data-efficient, easy to retrain, and avoids the instability and high overhead of RL.

\subsection{Considerations for Deployability and Adaptability}
Production schedulers must balance performance with operational simplicity. RL-based methods often require complex infrastructure for online training and are difficult to integrate into existing schedulers \cite{straesser2022not}. They also risk performance degradation during exploration phases. Heuristic systems are easier to deploy but require manual tuning and frequent intervention to remain effective under changing workloads \cite{augustyn2024tuning}.

Supervised learning offers a middle ground. Models can be trained offline on real telemetry data, validated against historical outcomes, and integrated into production with minimal runtime overhead. Retraining does not require system downtime or large-scale infrastructure changes.
\section{System Design} \label{sec:system_design}
We build a supervised learning–based scheduling framework for running network-intensive Spark jobs. The system is trained offline using historical job completion times, corresponding telemetry, and application configurations. Upon a job submission request, the framework ranks all candidate nodes using real-time network and node-level metrics, and selects the one predicted to offer the lowest job completion time. The framework runs externally in user space and integrates seamlessly with Kubernetes. Thus, it requires no modifications to Spark or Kubernetes control plane components.

\subsection{Background: Default Scheduling Behavior and Telemetry-Aware Decisions}
In Kubernetes, pod placement decisions are handled by the default scheduler (kube-scheduler), which assigns pods to nodes based on resource availability and policy constraints. This process involves two stages: filtering, where nodes that do not satisfy basic requirements (e.g., insufficient CPU/memory) are eliminated, and scoring, where remaining nodes are ranked using a set of scoring functions (e.g., least requested resources, affinity, taints/tolerations). The node with the highest score is then selected for placement \cite{senjab2023survey}.

However, the default scheduler is blind to runtime factors such as network variability, CPU pressure, or memory contention. It relies on static resource requests declared in pod specifications, rather than real-time system telemetry or application-specific performance considerations. As a result, even if a node satisfies the requested resources, actual job performance may degrade due to background load or interference from co-located pods. This is particularly problematic for data-intensive workloads, where performance is highly sensitive to such runtime variables.

To address the gap, telemetry-aware scheduling introduces live system signals into placement decisions. These signals may include network telemetry (e.g., inter-node RTTs, bandwidth usage), host-level metrics (e.g., CPU and memory pressure), or application-level traces. By leveraging such telemetry, a scheduler can better anticipate the performance impact of placement decisions and optimize for job-level objectives such as completion time.

\subsection{Architecture Overview} \label{sec:architecture_overview}
\begin{figure}
    \centering
    \includegraphics[width=\linewidth]{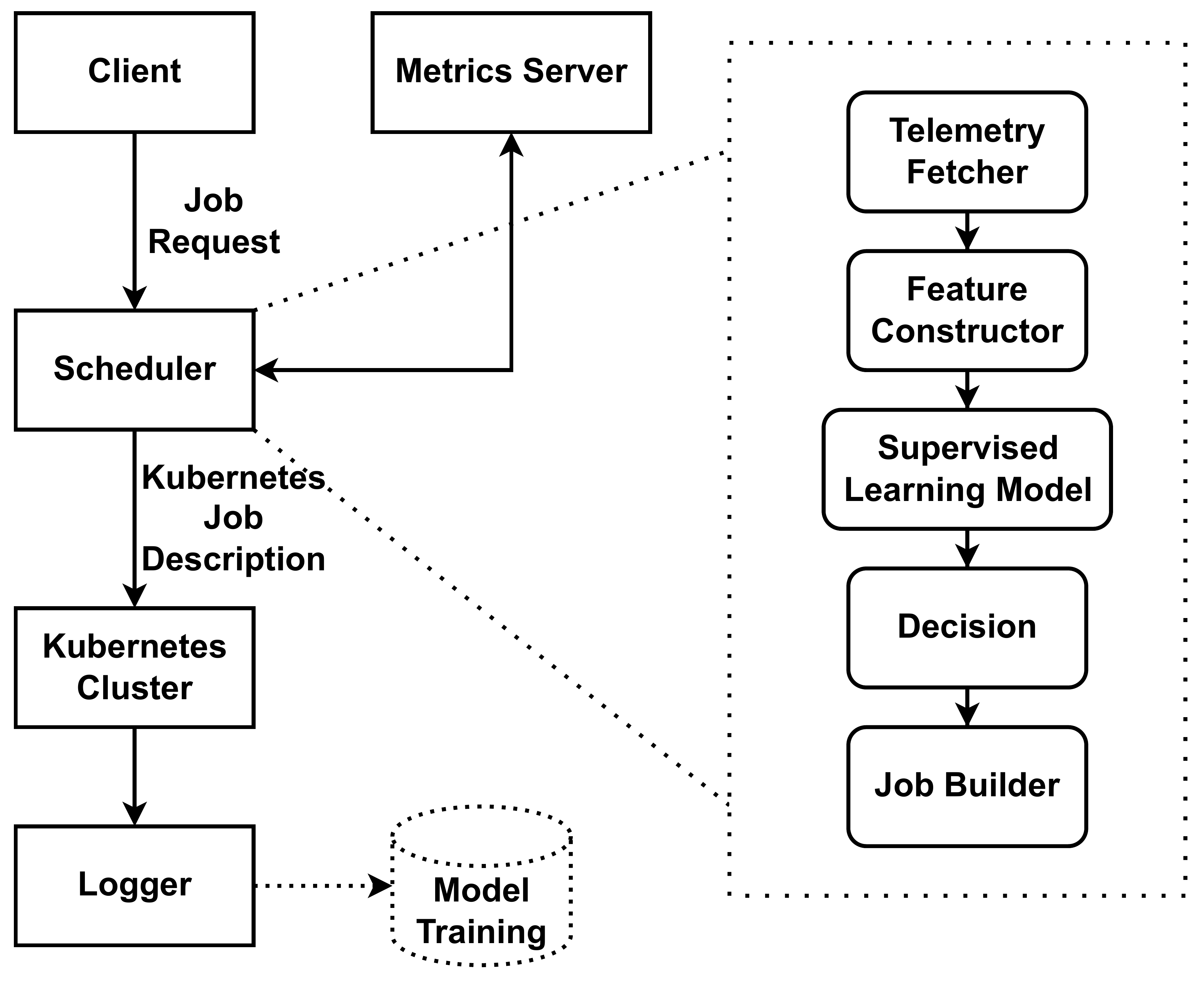}
    \caption{Overview of the system architecture.}
    \label{fig:architecture_overview}
\end{figure}

The overall architecture of the scheduling system is shown in Figure~\ref{fig:architecture_overview}. The system consists of five main components: (1) a client that issues job requests, (2) a metrics server that aggregates telemetry data across nodes, (3) a scheduler module that makes predictive placement decisions based on telemetry and a trained model, (4) a Kubernetes cluster that executes the jobs, and (5) a logger that stores outcomes for future retraining. In the following subsections, we describe each of these components in detail.

\subsubsection{Client}
The client is the entry point to the scheduling pipeline. It is responsible for initiating a job submission request, which includes application-specific parameters such as job type (e.g., sort, join), input data size, and resource configuration (e.g., executor count, memory). The request from the client is handled by the scheduler component (Section~\ref{sec:scheduler}) before it reaches the cluster. The client is agnostic to scheduling logic and only handles job initiation.

\subsubsection{Metrics Server} \label{sec:metrics-server}
The metrics server is responsible for aggregating and exposing system-level telemetry used by the scheduler for decision-making. In our system, this role is fulfilled by a Prometheus instance \cite{PrometheusGithub}, which is configured to scrape telemetry from multiple sources, including node-exporter for host-level statistics and custom ping mesh exporters for inter-node network latency.

The collected metrics provide a view of each node’s runtime state. These include: (a) Network-level telemetry: inter-node RTTs, transmit and receive rates, (b) Node-level telemetry: CPU load, and memory availability. The node-level metrics help capture transient runtime conditions that may not be visible from network telemetry alone.

\subsubsection{Scheduler} \label{sec:scheduler}
The scheduler is the decision-making component of the system. It receives job requests from the client, queries current telemetry from the metrics server, and selects the optimal node for job placement using a trained supervised learning model. The scheduler is implemented as an external user-space script that runs outside the Kubernetes control plane. To maintain modularity, the scheduler is internally composed of five submodules:

\paragraph{Telemetry Fetcher}
This component queries the Prometheus metrics server at scheduling time to retrieve the most recent telemetry snapshot. It fetches inter-node RTTs from the ping mesh, as well as per-node metrics such as CPU and memory load. These raw values are passed to the Feature Constructor for transformation.

\paragraph{Feature Constructor}
The feature constructor transforms raw telemetry and job configuration into structured input vectors for model inference. For each candidate node, it computes a fixed-size feature vector by combining telemetry data from the fetcher with static job-level attributes (e.g., input size, application type, memory requirement). These job-specific fields are critical because different jobs react differently to the same system state; for example, large shuffles are more sensitive to network load than CPU-bound tasks. Including this context helps the model make more accurate predictions across diverse workloads. The complete set of features is summarized in Table~\ref{tab:feature-set}.

\begin{table}[ht]
\caption{Input features used by the scheduling model.}
\label{tab:feature-set}
\centering
\normalsize
\begin{tabular}{|p{2.0cm}|p{4.0cm}|l|}
\hline
\textbf{Feature} & \textbf{Description} & \textbf{Type} \\
\hline
RTT            & Mean, max, and standard deviation of RTT to all peers & Network \\
Tx/Rx Rate     & Transmit and receive throughput (bytes/sec)           & Network \\
CPU            & CPU load average (runnable processes)                 & Node \\
Memory         & Available memory (bytes)                              & Node \\
Application Type & Categorical job type (e.g., sort, join)    & Job \\
Input Size     & Size of input data (e.g., number of records)          & Job \\
Other Configuration & Total executions, requested memory, etc. & Job \\
\hline
\end{tabular}
\end{table}

\paragraph{Supervised Learning Model}
The scheduler uses a supervised regression model to predict the expected job completion time for a given job on each candidate node. The model is trained offline using historical data collected from real job executions. Each training sample consists of a feature vector constructed from both node-level telemetry and job configuration features (Table~\ref{tab:feature-set}), along with the corresponding job completion time, which serves as the prediction target. We describe our application workloads and data characteristics in Section~\ref{sec:workloads}.

To evaluate modeling strategies, we experiment with several supervised learning approaches, including linear regression, random forests, and gradient-boosted decision trees (XGBoost)~\cite{scikitlearn,xgboost}. These models are well-suited to structured telemetry data due to their ability to capture nonlinear interactions and tolerate noisy or missing inputs, while maintaining fast inference and low deployment complexity. Random forests and XGBoost provide strong predictive accuracy and interpretable feature importance, while linear regression serves as a simple, interpretable baseline. We build and evaluate all three models.

\paragraph{Decision Module}
Once the supervised model predicts expected job completion times across candidate nodes, the scheduler ranks nodes in ascending order of predicted duration. The top-ranked node is selected as the launch node for the application.


\paragraph{Job Builder}
The module generates and submits jobs to the Kubernetes cluster based on the placement decision. It renders a declarative YAML manifest that is understood by Kubernetes for job launch. Node placement is enforced by injecting \texttt{nodeAffinity} rules into the generated specification. This ensures that the application is launched only on the selected node. Other parameters, such as job type, input size, and resource limits, are also populated dynamically at this stage.

\subsubsection{Kubernetes Cluster}
The system runs on a Kubernetes cluster deployed across geographically distributed sites on FABRIC. The nature of workloads is described in Section~\ref{sec:workloads}, and the cluster configuration is detailed in Section~\ref{sec:experimental_setup}.

\subsubsection{Logger}
The module captures telemetry and performance data at two stages of each job’s lifecycle. Before we submit a job, it records network and node-level telemetry (e.g., latency, throughput, CPU usage) to snapshot the system state. After the job completes, it collects application-level metrics such as job duration, stage execution times, and output size from the application logs. The collected data is used to support offline model training.
\section{Workloads} \label{sec:workloads}
To evaluate the effectiveness of our scheduler, we select a set of data-intensive applications representative of common patterns in distributed computing. Specifically, they differ in their reliance on shuffle-heavy operations (which involve large-scale data exchanges between executors during stages like grouping or sorting) and have varying degrees of inter-node communication and CPU utilization.

We evaluate our system using three Spark-based workloads~\cite{apachesparkdocs} with differing communication characteristics (Table~\ref{tab:workloads}). Each application launches a driver pod on a scheduler-selected node, while executor pods are placed independently by the default Kubernetes scheduler.

\begin{table}[ht]
\caption{Characteristics of the selected workloads.}
\label{tab:workloads}
\centering
\normalsize
\begin{tabular}{|l|p{6cm}|}
\hline
\textbf{Application} & \textbf{Rationale} \\
\hline
Sort & High network and CPU usage from large shuffles; moderate memory load \\
PageRank & High network and CPU usage from iterative data exchange; moderate memory load \\
Join & Skewed network, CPU, and memory usage due to imbalanced joins \\
\hline
\end{tabular}
\end{table}

\begin{figure}[ht]
    \centering
    \begin{minipage}{0.48\columnwidth}
        \centering
        \includegraphics[width=\linewidth]{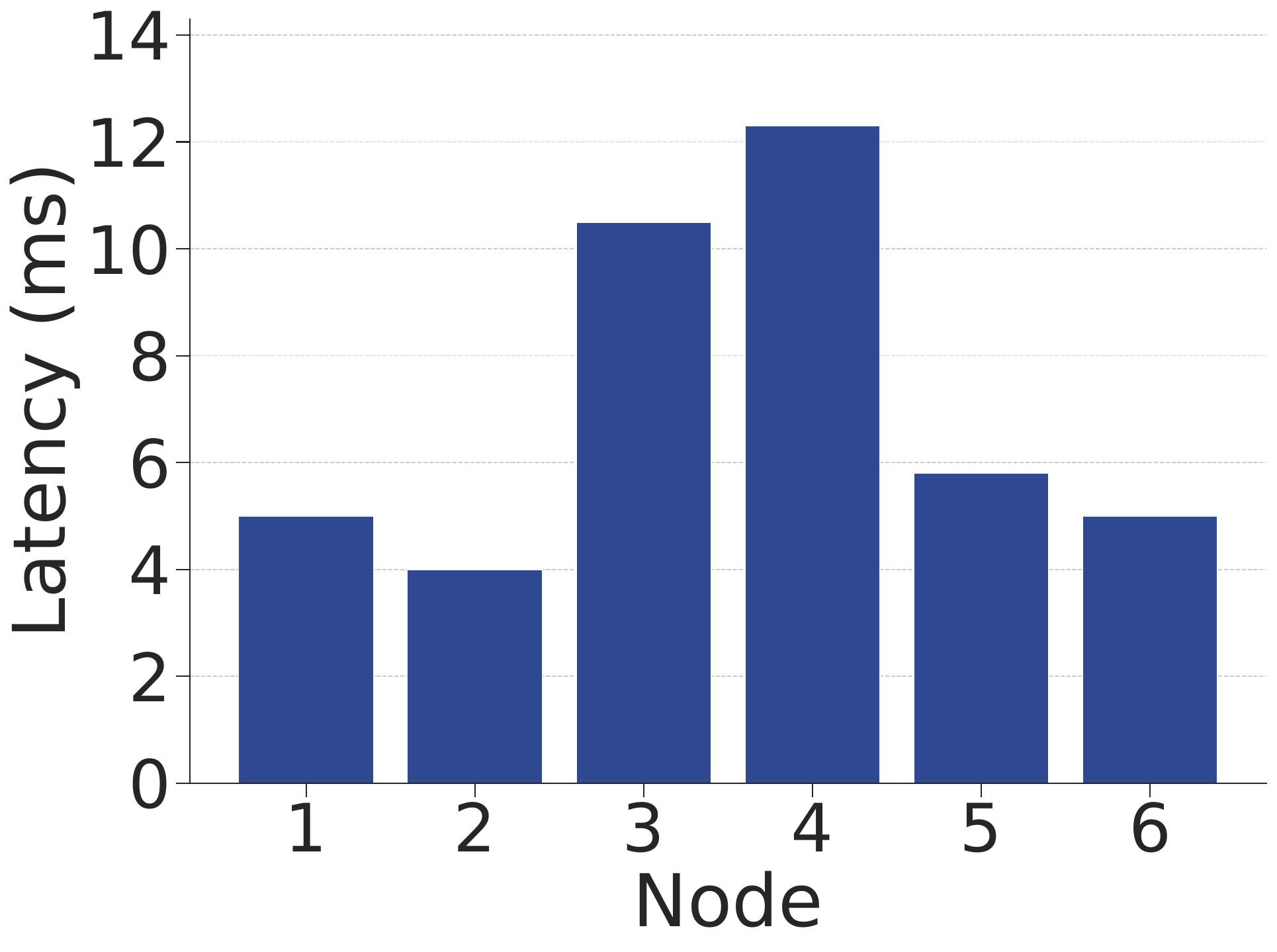}
        \caption{Average latency per node across five runs of \texttt{Sort}.}
        \Description{Average latency per node across five runs of \texttt{Sort}.}
        \label{fig:avg_latency}
    \end{minipage}
    \hfill
    \begin{minipage}{0.48\columnwidth}
        \centering
        \includegraphics[width=\linewidth]{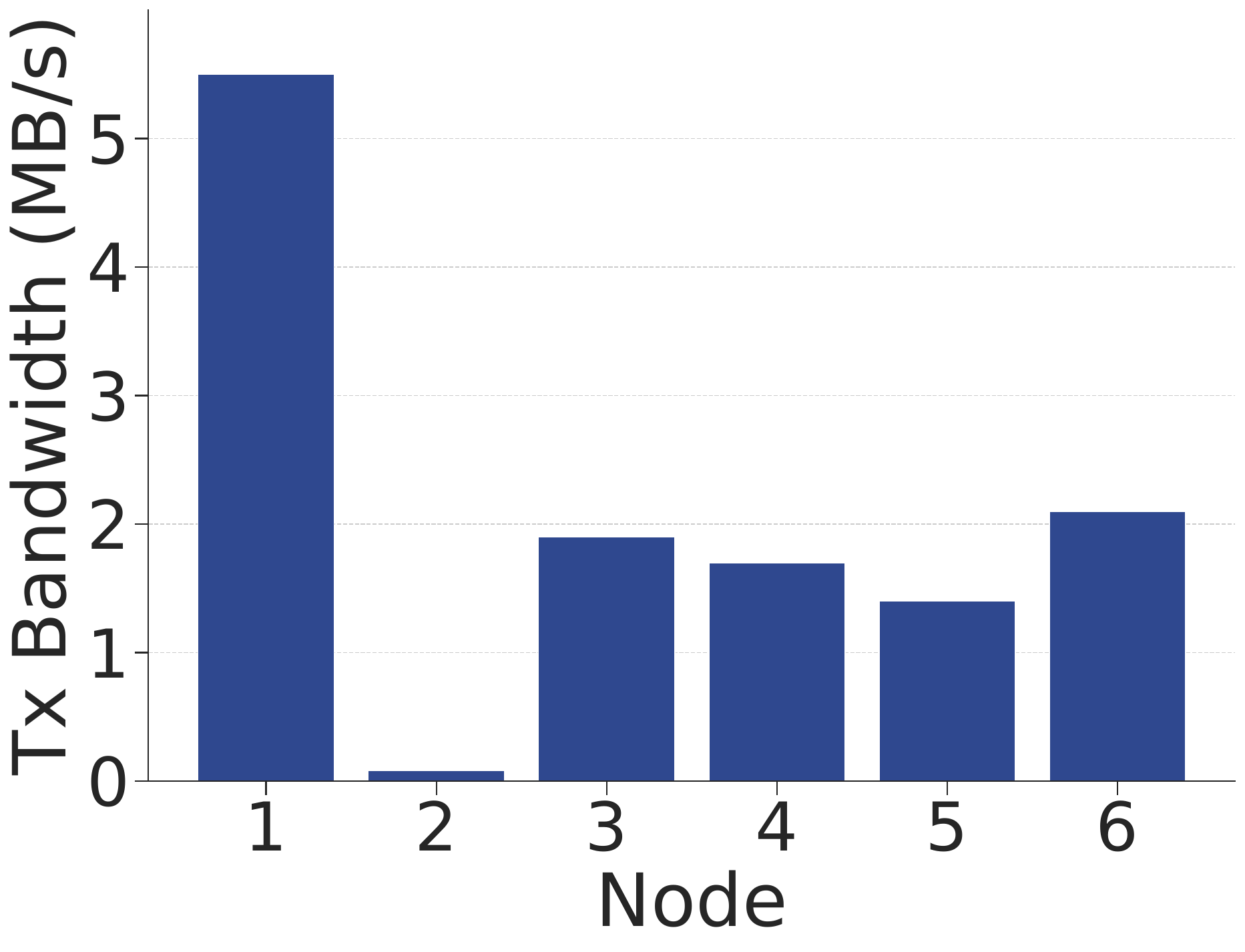}
        \caption{Average transmit bandwidth per node across five runs of \texttt{Sort}.}
        \Description{Average transmit bandwidth per node across five runs of \texttt{Sort}.}
        \label{fig:avg_tx}
    \end{minipage}
\end{figure}

The network telemetry of the \texttt{Sort} workload is shown in Figures~\ref{fig:avg_latency} and~\ref{fig:avg_tx}. These plots show differences in network latency and bandwidth across nodes. For example, \texttt{Node3} and \texttt{Node4} have higher latency, while \texttt{Node1} sees more transmit activity. The variation is captured during the learning phase and is fed to the scheduler.

\section{Experimental Setup} \label{sec:experimental_setup}
We evaluate our scheduler on a multi-site Kubernetes cluster deployed across geographically distributed nodes on the FABRIC testbed. This section describes the cluster configuration, and experiment workflow.

\subsection{Cluster Configuration}
We deploy a 6-node Kubernetes cluster across three FABRIC \cite{baldin2020fabric} sites: UC San Diego (UCSD), Florida International University (FIU), and SRI International (SRI), with two nodes provisioned at each site. The nodes are configured with 6~CPUs, 8\,GB RAM, and 25\,GB of disk space. Nodes are connected to isolated L3 IPv4 subnets using dedicated 100\,Gbps SR-IOV virtual NICs (Mellanox ConnectX-6). To enable inter-node communication, we configure static routes that forward traffic through the FABNetv4 data plane. FABNetv4 spans the FABRIC testbed and provides routing between these isolated subnets. This setup avoids using the management network for application traffic.

\begin{figure}
    \centering
    \includegraphics[width=\linewidth]{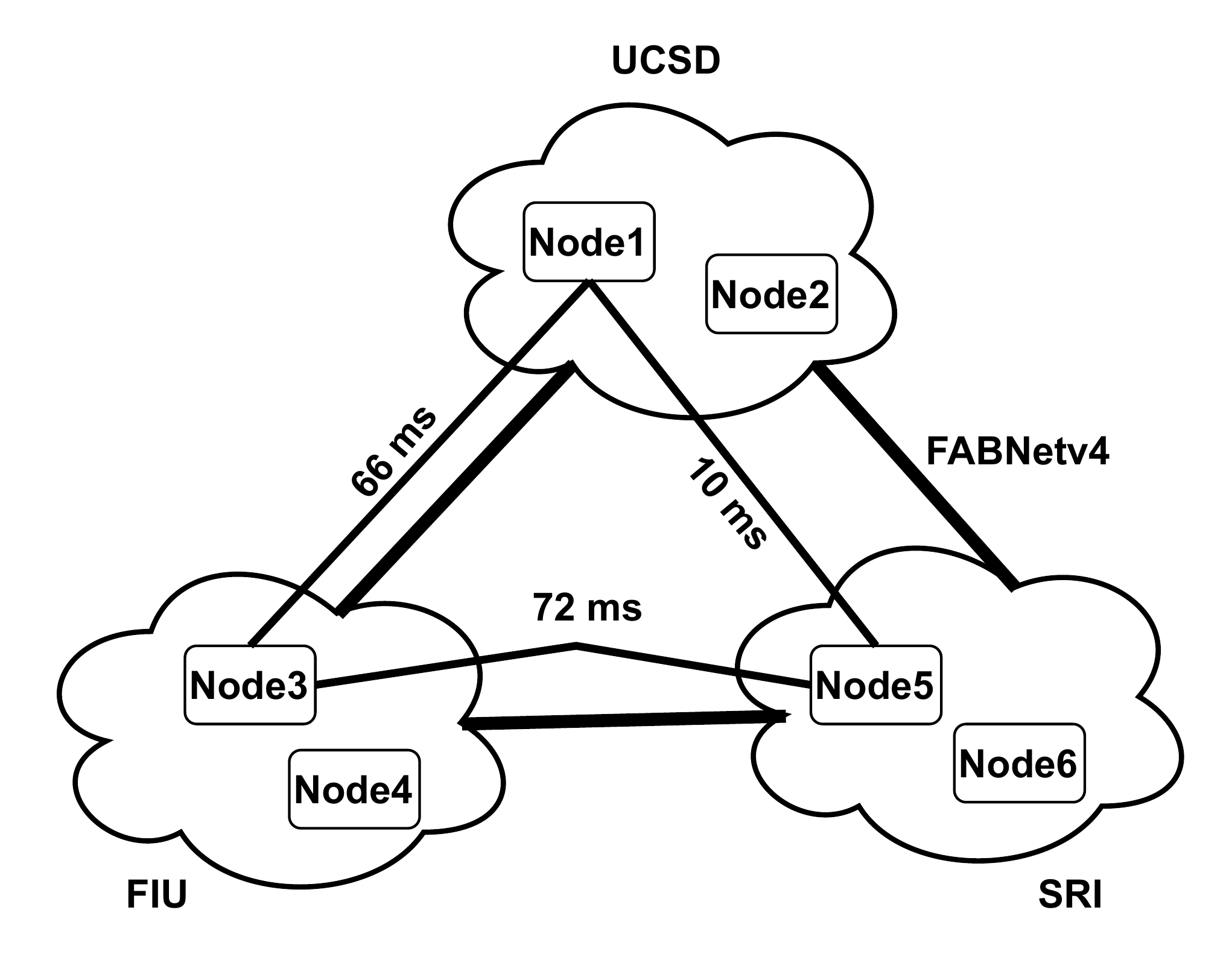}
    \caption{Geographical layout of the cluster across three FABRIC sites with RTT measurements shown along the connecting lines.}
    \Description{Geographical layout of the cluster across three FABRIC sites with RTT measurements shown along the connecting lines.}
    \label{fig:cluster_configuration}
\end{figure}

To run Spark applications, we use the Spark Operator. The operator allows submission and management of jobs through custom Kubernetes resources. For monitoring, we install the Prometheus stack to collect node-level telemetry for network, CPU, and memory usage. For inter-node RTT, we deploy \texttt{ping\_exporter}~\cite{pingexporter} as a DaemonSet. This enables full mesh probing across all nodes.

\subsection{Experiment Workflow}
We structure each experiment as a batch run and execute Spark jobs sequentially. Our setup includes 60 distinct job configurations across three Spark applications (Section~\ref{sec:workloads}) and covers a range of input sizes, executor counts, memory allocations, and shuffle patterns (e.g., group-by, join). To capture diverse runtime conditions, each configuration is executed across different \texttt{target\_node} values (e.g., node-1, node-6), and repeated 10 times. We introduced these variations because they affect both scheduling placement and telemetry input, and thus produce a broad range of system states. In total, our training dataset includes 3600 samples, where each sample represents a unique combination of job configuration and collected telemetry.

Background load (a pod that repeatedly downloads a 10MB file over HTTP using \texttt{curl}) is placed randomly on selected nodes during job execution. This simulates network and CPU contention, and leads to observable differences in metrics such as transmit/receive throughput and RTT. While simplified, this setup introduces variability closer to the dynamics observed in multi-tenant environments and helps the model learn from a broader range of system behaviors.

We train three supervised regression models using the full dataset: linear regression, random forest, and gradient-boosted decision trees (XGBoost). Each training sample is formed by joining node-level telemetry collected at scheduling time with static job configuration fields such as application type, input size, and resource requests (Table~\ref{tab:feature-set}). Telemetry is recorded before the job is launched because the model is designed to make proactive scheduling decisions. The job completion time obtained from application-level logs serves as the prediction target. A representative example of a training row is shown in Table~\ref{tab:training-row}.

\begin{table}[h]
\caption{Training sample: Input feature set (subset of the full feature set).}
\label{tab:training-row}
\begin{center}
\normalsize
\begin{tabular}{|p{1.0cm}|p{0.8cm}|p{0.8cm}|p{0.6cm}|p{0.6cm}|p{0.9cm}|p{0.8cm}|}
\hline
\textbf{RTT (s)} & \textbf{Rx (MB)} & \textbf{Tx (MB)} & \textbf{CPU (\%)} & \textbf{Mem (\%)} & \textbf{Input Size} & \textbf{Dur. (s)} \\
\hline
0.011 & 0.024 & 0.056 & 2.14 & 5.8 & 100000 & 18.18 \\
\hline
\end{tabular}
\end{center}
\end{table}
\section{Preliminary Results} \label{sec:preliminary_results}
We compared our scheduler against the Kubernetes default scheduling behavior and noted how accurately the scheduler’s choice matched the actual fastest node (Top-1) or included it among its two highest-ranked choices (Top-2). The actual fastest node was determined after the run based on the node that yielded the shortest job completion time. Random Forest achieved the highest Top-1 and Top-2 accuracies, followed by XGBoost and Linear Regression. Compared to the default scheduler, supervised learning approaches improved Top-1 accuracy by 34–54\% and Top-2 accuracy by 34–62\%, depending on the model.

\begin{table}[h]
\caption{Top-1 and Top-2 accuracy of different scheduling approaches in selecting the fastest execution node.}
\label{tab:prelim-results}
\begin{center}
\normalsize
\begin{tabular}{|p{3cm}|p{2.0cm}|p{2.0cm}|}
\hline
\textbf{Method} & \textbf{Top-1} & \textbf{Top-2} \\
\hline
Kubernetes Default   & 0.160 & 0.260 \\
Linear Regression    & 0.500 & 0.600 \\
XGBoost              & 0.560 & 0.720 \\
Random Forest        & \textbf{0.700} & \textbf{0.880} \\
\hline
\end{tabular}
\end{center}
\end{table}

While promising, these results are preliminary and drawn from a limited dataset in a controlled environment. They indicate the potential of supervised learning for network-aware scheduling, but further experimentation is needed to validate the findings under a wider range of workloads and cluster conditions (Section~\ref{sec:large_eval}).

\section{Related Work} \label{related_work}
Learning-based schedulers span several domains. DA-DRLS applies drift-adaptive deep reinforcement learning to IoT resource management, aiming to minimize energy consumption and response time under dynamic workloads \cite{chowdhury2019drls}. Marchese and Tomarchio extend the Kubernetes scheduler with a communication-aware plugin that models microservice dependencies, traffic history, and network latency to guide pod placement \cite{marchese2022communication}. Intel’s Telemetry-Aware Scheduling (TAS) \cite{intel_tas} is a scheduler extender that uses platform telemetry from the Custom Metrics API to apply policy-defined filtering, prioritization, and descheduling at placement time and during runtime. Zeus~\cite{zhang2021zeus} is a Kubernetes-compatible scheduler that colocates latency-sensitive services with best-effort jobs by monitoring server utilization, coordinating isolation features, and opportunistically placing workloads. It reports an increase in CPU utilization from 15\% to 60\% without SLO violations. In the Spark ecosystem, Shen et al. propose a multi-task prediction model that integrates convolutional layers and multi-head attention, applies dimensionality reduction to selected parameters, and estimates execution time across multiple applications \cite{shen2023novel}. For ML inference clusters, Eddine et al. propose a tail-latency-aware scheduler for real-time workloads on resource-constrained devices \cite{eddine2025tail}. Their framework combines AI model features and hardware profiles, using offline-trained regression models and an online scoring function to guide task placement. Beyond scheduling and job prediction, prior systems have explored network-centric mechanisms to improve performance and flexibility in distributed environments. N-DISE~\cite{wu2022n} achieved sustained \textgreater31 Gbps WAN throughput in CMS workflows by exploiting in-network caching, while LIDC~\cite{timilsina2024lidc} introduced a location-independent framework that routes compute requests by name via in-network routers using Kubernetes.

\section{Conclusion and Future Work} \label{sec:conclusion_and_future_work}
This paper presented a network-aware scheduler that uses supervised learning to predict job completion time based on real-time telemetry. Through initial experiments on shuffle-heavy Spark workloads across a geographically distributed Kubernetes cluster, we showed that the scheduler can outperform the Kubernetes default scheduling behavior. In particular, the supervised learning model enabled more informed placement decisions, and placed jobs on better-performing nodes with accuracy gains of 34-54\% over the default scheduler. We outline several technical directions to expand and refine our system:

\paragraph{Evaluation at a larger scale with real workloads} \label{sec:large_eval}
While our current evaluation focused on a limited dataset and representative jobs, we plan to deploy the scheduler on larger topologies with real-world, data-intensive applications such as distributed ML pipelines, iterative graph algorithms, and multi-stage streaming jobs. This will allow us to validate the scheduler’s performance and scalability under production-like conditions and a wider range of workload characteristics.

\paragraph{Quantifying deployability and retraining costs} 
Our system is designed to be modular, lightweight and easy to integrate across diverse environments, but further evaluation is needed to measure the overhead of telemetry collection, model retraining frequency, and the trade-offs between model accuracy and scheduling latency. Understanding these costs will help assess how practical the system is for dynamic or production workloads.

\paragraph{Richer network telemetry integration} 
We currently rely on node-level metrics such as RTT and aggregate bandwidth usage. We aim to incorporate fine-grained telemetry such as link-level utilization (e.g., per-interface throughput and congestion), queuing delay (e.g., estimated buffer occupancy), and passive flow-level statistics (e.g., flow count, duration). These metrics can help capture transient congestion, path-level variability, and network bottlenecks that are not visible at the node level.

\begin{acks}
This material is based upon work supported by the National
Science Foundation under grants 2126148 and 2430341.    
\end{acks}

\bibliographystyle{ACM-Reference-Format}
\bibliography{references}

\end{document}